\documentclass[pra,aps,nopacs,twocolumn,twoside,superscriptaddress]{revtex4}

\usepackage{amsmath,amsfonts,amssymb,color,graphics,graphicx,subfigure,latexsym,revsymb,amsthm,url,verbatim,appendix,epstopdf,floatrow,booktabs}

\usepackage{hyperref}

\usepackage{tikz}
\usetikzlibrary{decorations.shapes}
\usetikzlibrary{shapes.symbols}

\newtheorem{theorem}{Theorem}
\newtheorem{corollary}[theorem]{Corollary}
\newtheorem{lemma}[theorem]{Lemma}

\theoremstyle{definition}
\newtheorem{definition}{Definition}
\newtheorem{proposition}[definition]{Proposition}

\newcommand{\meas}{
\begin{tikzpicture}
\filldraw[fill=white] (0,.25) rectangle (.7,-.25);
\draw (.67,-.1) arc (50:130:.5);
\draw (.35,-.2)--(.525,.2);
\end{tikzpicture}
}

\newcommand{\cntrl}{
\begin{tikzpicture}
\fill (0,0) circle (.08);
\end{tikzpicture}
}

\def\squareforqed{\hbox{\rlap{$\sqcap$}$\sqcup$}}
\def\qed{\ifmmode\squareforqed\else{\unskip\nobreak\hfil
\penalty50\hskip1em\null\nobreak\hfil\squareforqed
\parfillskip=0pt\finalhyphendemerits=0\endgraf}\fi}
\def\endenv{\ifmmode\;\else{\unskip\nobreak\hfil
\penalty50\hskip1em\null\nobreak\hfil\;
\parfillskip=0pt\finalhyphendemerits=0\endgraf}\fi}
\def\Dbar{\leavevmode\lower.6ex\hbox to 0pt
{\hskip-.23ex\accent"16\hss}D}

\makeatletter
\def\bcj{\begin{conjecture}}
\def\ecj{\end{conjecture}}
\def\bcr{\begin{corollary}}
\def\ecr{\end{corollary}}
\def\bd{\begin{definition}}
\def\ed{\end{definition}}
\def\bea{\begin{eqnarray}}
\def\eea{\end{eqnarray}}
\def\bem{\begin{enumerate}}
\def\eem{\end{enumerate}}
\def\bex{\begin{example}}
\def\eex{\end{example}}
\def\bim{\begin{itemize}}
\def\eim{\end{itemize}}
\def\bl{\begin{lemma}}
\def\el{\end{lemma}}
\def\bpf{\begin{proof}}
\def\epf{\end{proof}}
\def\bpp{\begin{proposition}}
\def\epp{\end{proposition}}
\def\bqu{\begin{question}}
\def\equ{\end{question}}
\def\br{\begin{remark}}
\def\er{\end{remark}}
\def\bt{\begin{theorem}}
\def\et{\end{theorem}}

\def\btb{\begin{tabular}}
\def\etb{\end{tabular}}

\newcommand{\nc}{\newcommand}

\def\G{\Gamma}

 \nc{\bA}{{\bf A}} \nc{\bB}{{\bf B}} \nc{\bC}{{\bf C}}
 \nc{\bD}{{\bf D}} \nc{\bE}{{\bf E}} \nc{\bF}{{\bf F}}
 \nc{\bG}{{\bf G}} \nc{\bH}{{\bf H}} \nc{\bI}{{\bf I}}
 \nc{\bJ}{{\bf J}} \nc{\bK}{{\bf K}} \nc{\bL}{{\bf L}}
 \nc{\bM}{{\bf M}} \nc{\bN}{{\bf N}} \nc{\bO}{{\bf O}}
 \nc{\bP}{{\bf P}} \nc{\bQ}{{\bf Q}} \nc{\bR}{{\bf R}}
 \nc{\bS}{{\bf S}} \nc{\bT}{{\bf T}} \nc{\bU}{{\bf U}}
 \nc{\bV}{{\bf V}} \nc{\bW}{{\bf W}} \nc{\bX}{{\bf X}}
 \nc{\bZ}{{\bf Z}}

\nc{\cA}{{\cal A}} \nc{\cB}{{\cal B}} \nc{\cC}{{\cal C}}
\nc{\cD}{{\cal D}} \nc{\cE}{{\cal E}} \nc{\cF}{{\cal F}}
\nc{\cG}{{\cal G}} \nc{\cH}{{\cal H}} \nc{\cI}{{\cal I}}
\nc{\cJ}{{\cal J}} \nc{\cK}{{\cal K}} \nc{\cL}{{\cal L}}
\nc{\cM}{{\cal M}} \nc{\cN}{{\cal N}} \nc{\cO}{{\cal O}}
\nc{\cP}{{\cal P}} \nc{\cQ}{{\cal Q}} \nc{\cR}{{\cal R}}
\nc{\cS}{{\cal S}} \nc{\cT}{{\cal T}} \nc{\cU}{{\cal U}}
\nc{\cV}{{\cal V}} \nc{\cW}{{\cal W}} \nc{\cX}{{\cal X}}
\nc{\cZ}{{\cal Z}}

\nc{\hA}{{\hat{A}}} \nc{\hB}{{\hat{B}}} \nc{\hC}{{\hat{C}}}
\nc{\hD}{{\hat{D}}} \nc{\hE}{{\hat{E}}} \nc{\hF}{{\hat{F}}}
\nc{\hG}{{\hat{G}}} \nc{\hH}{{\hat{H}}} \nc{\hI}{{\hat{I}}}
\nc{\hJ}{{\hat{J}}} \nc{\hK}{{\hat{K}}} \nc{\hL}{{\hat{L}}}
\nc{\hM}{{\hat{M}}} \nc{\hN}{{\hat{N}}} \nc{\hO}{{\hat{O}}}
\nc{\hP}{{\hat{P}}} \nc{\hR}{{\hat{R}}} \nc{\hS}{{\hat{S}}}
\nc{\hT}{{\hat{T}}} \nc{\hU}{{\hat{U}}} \nc{\hV}{{\hat{V}}}
\nc{\hW}{{\hat{W}}} \nc{\hX}{{\hat{X}}} \nc{\hZ}{{\hat{Z}}}

\nc{\hn}{{\hat{n}}}

\def\diag{\mathop{\rm diag}}

\def\ox{\otimes}

\newcommand{\ket}[1]{|#1\rangle}

\newcommand{\tgate}{{\sf T}}
\newcommand{\hgate}{{\sf H}}
\newcommand{\pgate}{{\sf P}}
\newcommand{\igate}{{\mathbb I}}
\newcommand{\cnot}{{\sf CNOT}}

\newcommand{\Xgate}{{\sf X}}
\newcommand{\Zgate}{{\sf Z}}
\newcommand{\Ygate}{{\sf Y}}

\begin{document}
\title{A framework for quantum homomorphic encryption with experimental demonstration}
\author{Yu Zhang}
\affiliation{Department of Physics, Hangzhou Normal University, Hangzhou, Zhejiang 311121, China}
\affiliation{School of Physics, Nanjing University, Nanjing, Jiangsu 210093, China}
\author{Li Yu}\email{yupapers@sina.com}
\author{Qi-Ping Su}
\author{Zhe Sun}
\affiliation{Department of Physics, Hangzhou Normal University, Hangzhou, Zhejiang 311121, China}
\author{Fuqun Wang}
\affiliation{Department of Mathematics, Hangzhou Normal University, Hangzhou, Zhejiang 311121, China}
\affiliation{Westone Cryptologic Research Center, Beijing 100071, China}
\author{Xiao-Qiang Xu}
\author{Qingjun Xu}
\author{Jin-Shuang Jin}
\affiliation{Department of Physics, Hangzhou Normal University, Hangzhou, Zhejiang 311121, China}
\author{Kefei Chen}
\affiliation{Department of Mathematics, Hangzhou Normal University, Hangzhou, Zhejiang 311121, China}
\affiliation{Westone Cryptologic Research Center, Beijing 100071, China}
\author{Chui-Ping Yang}\email{yangcp@hznu.edu.cn}
\affiliation{Department of Physics, Hangzhou Normal University, Hangzhou, Zhejiang 311121, China}

\date{\today}

\begin{abstract}
Quantum homomorphic encryption (QHE) is an encryption method that allows quantum computation to be performed on one party's private data with the program provided by another party, without revealing much information about the data nor the program to the opposite party. We propose a framework for (interactive) QHE based on the universal circuit approach. It contains a subprocedure of calculating a classical linear polynomial, which can be implemented with quantum or classical methods; apart from the subprocedure, the framework has low requirement on the quantum capabilities of the party who provides the circuit. We illustrate the subprocedure using a quite simple classical protocol with some privacy tradeoff. For a special case of such protocol, we obtain a scheme similar to blind quantum computation but with the output on a different party. Another way of implementing the subprocedure is to use a recently studied quantum check-based protocol, which has low requirement on the quantum capabilities of both parties. The subprocedure could also be implemented with a classical additive homomorphic encryption scheme. We demonstrate some key steps of the outer part of the framework in a quantum optics experiment.
\end{abstract}
\maketitle


\section{Introduction}\label{sec1}

Secure quantum computing is an interesting research field that draws ideas from classical cryptography and quantum mechanics. It aims to keep the input data or the program private when two parties perform a quantum computation on possibly restricted sets of input states, while the input and the output may be on one or both parties. Blind quantum computing (BQC) \cite{bfk09,ABE10,skm13,Barz12} and  ``quantum computing on encrypted data'' \cite{Ch05,Fisher13} are two types of problems for which there are satisfactory protocols. Quantum homomorphic encryption is a different but related problem, and is the topic of this paper.

Blind quantum computing is a task in which a client knows both the data and the program but can only do limited quantum operations, and the main part of the computation is carried out by the server, who is not able to learn the data nor the program by design of the protocol. The BQC scheme in \cite{bfk09} builds on the measurement-based quantum computing model. There are other BQC schemes based on the circuit model \cite{ABE10}, or the ancilla-driven model \cite{skm13}. On the other hand, ``Quantum computing on encrypted data'', or delegated quantum computation \cite{Ch05,Fisher13}, allows a publicly known quantum program to be run by a server on secret quantum data provided by a client.

In classical cryptography, homomorphic encryption (HE) is an encryption scheme that allows computation to be performed (effectively on the plaintext after decryption) while having access only to the ciphertext. A \emph{fully} homomorphic encryption (FHE) scheme allows for \emph{any} computation to be performed in such fashion \cite{Gentry09,Dijk09,smart2010fully,brakerski2011efficient}. Quantum homomorphic encryption (QHE) allows some classes of quantum computation to be performed without accessing the unencrypted data, and quantum fully homomorphic encryption (QFHE) is similarly defined with the allowed class of quantum computation being universal. The goals of QHE are that the final computation result is correct, and the data and the final computation result are known only to the data-provider, who learns little about the circuit performed beyond what can be deduced from the computation result itself. Schemes for QHE \cite{rfg12,MinL13,Tan16,Ouyang18,bj15,Dulek16,NS17,Lai17,ADSS17,Mahadev17,TOR18} with varying degrees of success have been proposed. Some techniques in QHE have been applied to the problem of computing on shared quantum secrets \cite{Ouyang17}. QHE schemes for arbitrary unrestricted circuits are shown to require exponential resources \cite{ypf14,NS17,Lai17,Newman18}. Although \cite{ypf14} and some results on bipartite quantum computation of classical functions \cite{Lo97,bcs12,Colbeck07} suggest that the goal of both perfect data privacy and near-perfect circuit privacy cannot be reached in the plain model without aborts, the recent work \cite{Yu19} suggests that when some party is allowed to abort, some nontrivial level of security can be reached. An experimental implementation of homomorphic-encrypted quantum walk proposed in \cite{rfg12} has been reported \cite{Zeuner18}. An experimental demonstration of computationally-secure QFHE is shown in \cite{Tham18}.

In this paper, we introduce a framework for (interactive) QHE schemes, intended to work for general quantum input with a polynomial-sized circuit. We refer to ``interactive QHE scheme'' as any protocol that achieves the goals of QHE (the correctness of final result, and the security of the data and the circuit) while allowing multiple rounds of communication. The flexibility in the framework is that one subprocedure, namely the evaluation of the classical linear polynomials, can be replaced by any quantum or classical protocol for such task. The main computation is on the side of Alice, who provides the data. The framework depends on a ``program state'' method which is seen in ancilla-driven blind quantum computing \cite{skm13}, and it takes hint from a ``universal circuit'' method in \cite[Supplementary Note 1]{Fisher13}. If not considering the implementation of the subprocedure, the party Bob who provides the circuit need only have very limited quantum capability: to prepare and send single-qubit states in one of two bases, or to make single-qubit measurements in one of two bases. This point distinguishes it from the framework in \cite{bj15,Dulek16}, where the main quantum computation is carried out by the party who provides the circuit, so that both parties need to be able to store at least $n$ qubits (in the case of quantum input; and $n$ is the input size).

We provide two example classes of protocols for the subprocedure for evaluating the classical linear polynomials. The first class is entirely classical, and it satisfies that the degree of data privacy is in a tradeoff with that of the circuit privacy, both in the subprocedure and in the overall scheme. One end of such tradeoff is that the data is completely insecure, and then the circuit is quite (but not completely) secure, which yields a scheme similar to blind quantum computing, but the output is on the Alice (server) side, rather than on the party with the initial program. When the input state is classical or a product real state, the tradeoff between the data privacy and the circuit privacy is unexpectedly good compared to the case of general quantum input. The second way of implementing the subprocedure is to use a recently studied quantum check-based protocol \cite{Yu19}, which has low requirement on the quantum capabilities of both parties: each party need only do repeated quantum operations on a constant number of qubits at a time.

The framework under the two classes of protocols for evaluating classical linear polynomials have modest resource costs: for the first class, the entanglement and classical communication costs are linear in the product of circuit size and the input size; for the second class, the costs are polynomial in circuit size and the input size.

We demonstrate the key steps of the outer part of framework (but not the subprocedure for evaluating classical linear polynomials) in a quantum optics experiment.

In Sec.~\ref{sec4}, we mention that the subprocedure for evaluating the classical linear polynomials could also be implemented with a classical additive homomorphic encryption scheme.

The rest of the paper is organized as follows. Sec.~\ref{sec2} presents the framework for interactive QHE, and briefly introduces the two classes of protocols for the subprocedure of evaluation of the classical linear polynomials. Sec.~\ref{sec3} presents the experimental methods and the results of the experiment. Sec.~\ref{sec4} contains some discussions. Sec.~\ref{sec5} contains the conclusions.

\section{Framework for interactive QHE}\label{sec2}

First, we introduce the method of using program states in our scheme. A circuit diagram is shown in Fig.~\ref{fig1} below. The lower qubit is referred to as the \emph{program register}, and its state $\phi(\theta)=\frac{1}{\sqrt{2}}(\cos\theta\ket{0}+i\sin\theta\ket{1})$ is called the \emph{program state}. It interacts with a data qubit (the first qubit in the figure) via a fixed controlled-$\sigma_j$ gate (where $j\in\{1,2,3\}$, and $\sigma_1=\sigma_x$, $\sigma_2=\sigma_y$, $\sigma_3=\sigma_z$ are the Pauli operators), and then the program register is subject to a Hadamard gate $\hgate$, and measured in the $Z$ basis. (By viewing the Hadamard gate as a change of basis, the lower qubit is effectively measured in the $X$ basis.) When the measurement result is $0$, the one-qubit gate $\cos(\theta)\igate+i\sin(\theta)\sigma_j$ is implemented on the first qubit. When the measurement result is $1$ and the program state is not in the $Z$ basis (i.e. $\theta\ne\frac{k\pi}{2}$ for integer $k$), a unitary correction is needed, to make the effective gate on the data qubit the same as in the case that the measurement outcome is $0$. The form of the correction depends on the program state: it is $\cos (2\theta)\,\igate+i\sin (2\theta)\,\sigma_j$. When the correction is a Pauli operator not equal to $\igate$, which happens when $\theta=\frac{k\pi}{4}$ with odd $k$, this correction can be deferred by commuting it through the Clifford gates in later parts of the protocol. Note that two mutually orthogonal states of the program register correspond to the same program up to a Pauli correction known to Bob. If Alice knows that the program state is one of these two states, she can of course make a projective measurement to distinguish them. But when Alice is not sure that the program state is one of these two states, e.g. when there are four possible states in two different bases, she cannot know the exact program state, which happens in the case discussed below Eq.~\eqref{eq:Ggates}.

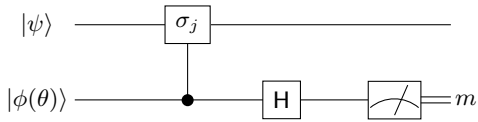
\begin{figure}[h]
\centering
\begin{tikzpicture}
\node at (-0.5,5) {$\ket{\psi}$};
\draw (0,5)--(5,5);
\node at (-0.5,4) {$\ket{\phi(\theta)}$};
\draw (0,4)--(4.25,4);

\filldraw[fill=white] (1.2,5.25) rectangle (1.8,4.75);
\node at (1.5,5) {$\sigma_j$};
\node at (1.5,4) {\cntrl};
\draw (1.5,4.75)--(1.5,4);

\filldraw[fill=white] (2.5,4.25) rectangle (3.0,3.75);
\node at (2.75,4) {$\hgate$};

\node at (4.25,4) {\meas};
\draw (4.6,4.04)--(5,4.04);
\draw (4.6,3.96)--(5,3.96);
\node at (5.2,4) {$m$};

\end{tikzpicture}
\caption{The gadget for using the program state (in the lower qubit) to implement a gate $\cos(\theta)\igate+i\sin(\theta)\sigma_j$ on the data qubit (the upper qubit). This circuit is done entirely by Alice in Scheme 1. The lower qubit is initially in a real state $\phi(\theta)=\frac{1}{\sqrt{2}}(\cos\theta\ket{0}+i\sin\theta\ket{1})$. The two-qubit gate is controlled-$\sigma_j$, where $j\in\{1,2,3\}$. By viewing the Hadamard gate as a change of basis, the lower qubit is effectively measured in the $X$ basis. For the measurement outcome $m=1$, a unitary correction $\cos (2\theta)\,\igate+i\sin (2\theta)\,\sigma_j$ is required on the data qubit. But in Scheme 1, rather than doing the correction immediately, Bob updates his coefficients in some linear polynomials of the form \eqref{eq:poly} in order to do corrections later.}\label{fig1}
\end{figure}

In the scheme below, some program states are produced by Bob through his measurement on some shared entangled state with Alice. From a security point of view, such remote operation is not really necessary. Bob could directly send some states to Alice as the program states, or directly prepare some program states using Alice's experimental apparatus (in the latter case, for security of Alice's data, we have to assume that Bob has no access to Alice's data states when doing so). Alice encrypts her data using the quantum one-time-pad first, before letting her data interact with the program states. Such encryption is for protecting the data from possible leakage due to that some measurement outcomes are sent to Bob. See Sec.~\ref{sec4} for more explanations.

Since every one-qubit Clifford gate can be expressed as the product of some Clifford gates of the form
\bea\label{eq:onequbitCL}
&&\cos(\theta)\igate+i\sin(\theta)\sigma_j,\notag\\
&&\mbox{with}\quad \theta=\frac{k\pi}{4},\quad\mbox{for}\quad k\in\{0,1,2,3\},
\eea
the program states can be used to implement one-qubit Clifford gates. Alice can of course implement two-qubit gates by herself, but we need a nontrivial two-qubit gate or the identity to be implemented according to Bob's choice. This is done as follows. Let $\G=\frac{1-i}{2}(\igate\ox\igate+i \Zgate\ox\Zgate)=\diag\left(1,-i,-i,1\right)$. It is a two-qubit Clifford gate. The following construction accommodates a free choice of a gate in the set $\{\igate\ox\igate,\G\}$ using a fixed sequence of $\G$ interleaved with single-qubit Clifford gates:
\begin{eqnarray}\label{eq:Ggates}
\G \left( \igate\ox\hgate \right) \G \left( \igate\ox\hgate \right) \G \left( \Zgate\ox\hgate\right) \G&=&e^{-\pi i/4}\G, \notag\\
\G\cdot\G\cdot\G\cdot\G&=&\igate\ox\igate.
\end{eqnarray}
The $\G$ gate can be implemented by Alice. The one-qubit Clifford gates in Eq.~\eqref{eq:Ggates} can be implemented by the program-state method, and Bob has a choice as to whether to implement the $\hgate$ or $\igate$ by choosing appropriate program states, according to whether he wants to implement the overall gate $\G$ or $\igate\ox\igate$. The $\hgate$ and $\igate$ are both products of some one-qubit Clifford gates of the form \eqref{eq:onequbitCL} that are implementable by program states. (The Pauli corrections in the implementations can be delayed.) Thus the choice between $\hgate$ and $\igate$ is realized by different choices of program states at each gate of the form \eqref{eq:onequbitCL}. For each one-qubit Clifford gate of the form \eqref{eq:onequbitCL}, the average state of the two possible program states is the maximally mixed state. Hence, Alice is ignorant of which gate is implemented, when she has not received any information about the Pauli corrections. (Another way to see this fact is to consider the generation of program states by Bob's measurement on his qubit in an EPR pair, which is in Scheme 1 below. Before Bob tells Alice any information about the Pauli corrections, Alice received no messages, so she had no information about which gate Bob intends to implement.)

Alice initially encrypts her input data qubits using the quantum one-time-pad, which means applying random Pauli operators. The choice of the Pauli operators are recorded as $2n$ bits: the Pauli operator $\Xgate^j \Zgate^k$ is recorded as two bits $j$ and $k$. These $2n$ bits are unknown to Bob, so he regards them as variables. The variables are unchanged in the procedures of the scheme below, and Bob changes the coefficients for them in the polynomials in the scheme [i.e. in Eq.~\eqref{eq:poly}, Bob updates the coefficients $a_i$ but not the variables $x_i$]. The way Bob changes the coefficients is because of some key-update rules, and we call them effective key-update rules below, since they do not change Alice's keys but rather change Bob's coefficients. The effective key-update rules can be easily obtained from the following relations:
\begin{eqnarray}\label{eq:keyupdate1}
&&\pgate\Xgate=i\Xgate\Zgate\pgate,\quad\quad \pgate\Zgate=\Zgate\pgate,\notag\\
&&\hgate\Xgate=\Zgate\hgate,\quad\quad\quad \hgate\Zgate=\Xgate\hgate,\notag\\
&&\cnot_{12} (\Xgate_1^a \Zgate_1^b \ox \Xgate_2^c \Zgate_2^d)= (\Xgate_1^a \Zgate_1^{b\oplus d} \ox \Xgate_2^{a\oplus c} \Zgate_2^d)\cnot_{12},\notag\\
&&
\end{eqnarray}
where the $\oplus$ is addition modulo 2, and in the gate $\cnot_{12}$, the qubit 1 is the control. The effective key-update rules under the $\tgate$ gate can be obtained from the relations
\begin{eqnarray}\label{eq:keyupdate2}
\tgate\Zgate=\Zgate\tgate,\quad\quad \tgate\Xgate=e^{-\pi i/4}\pgate\Xgate\Zgate\tgate.
\end{eqnarray}
More details about the key-update rules are in \cite{bj15,Dulek16}.

The detailed steps of our scheme are shown in Scheme 1 below. Starting from step 2, we distinguish the case of classical or product real input from the case of general quantum input. For each $\tgate$ gate in the desired circuit, Alice and Bob perform a subprocedure to evaluate a classical linear polynomial of the form
\bea\label{eq:poly}
y=(c+\sum_{i=1}^{m} a_i x_i) \mod 2,
\eea
with the output being a bit on Alice's side, indicating whether a $\pgate^\dag$ gate should be done after the $\tgate$ gate, to compensate for differences on different Pauli masks due to the $\tgate$ gate. The $a_i$ are \emph{coefficients} known to Bob, and the $x_i$ are \emph{variables} known to Alice. The $m$ is $2n$ in the case of general quantum input, but is $n$ in the case of classical or product real input, since then the initial Pauli corrections are only of the $Y=\sigma_y$ type, which suffice for making the average state maximally mixed. The ``classical input'' refers to that the input qubits are in the state $\ket{0}$ or $\ket{1}$. The ``product real input'' refers to that the input is the tensor product of real qubit states. For the stage of Clifford gates after the last $\tgate$ gate, Alice and Bob perform two instances of the subprocedure for each of the output qubits, and the output bits are for what Pauli corrections Alice should do on each output qubit.

{\bf Scheme 1.} \emph{A framework for interactive QHE.}

1.  Bob decomposes the desired circuit using Clifford and $\tgate$ gates, where the Clifford gates involved only include the $\G$ gate and the single-qubit Clifford gates. Alice and Bob prepare some EPR pairs in the state $(\ket{00}+\ket{11})/\sqrt{2}$. Bob measures his qubits in the entangled pairs in the $Z$ or $Y$ basis, according to the desired program to be performed on Alice's data. After the measurement, Alice's qubit in each pair (denoted $a$) is in one of the following four states in Bob's view: $\ket{0}$, $\ket{1}$, $\ket{y+}$, $\ket{y-}$, which are referred to as the \emph{program states}. In this way, Bob effectively sends the program states to Alice.

2. In the case of general quantum input, Alice does a random Pauli gate on each input data qubit, and locally records the $2n$ Pauli mask bits, with bit value $1$ indicating a Pauli gate $X$ or $Z$ was done. These $2n$ bits are Alice's variables in the linear polynomials to be evaluated by both parties. In the case that the input is classical or a direct product of real qubit states, Alice does a random Pauli $\Ygate=\sigma_y$ gate on each input data qubit, and locally records the $n$ Pauli mask bits, with bit value $1$ indicating a Pauli gate $\Ygate$ was done.

3. Bob sends Alice some message about which qubit each program state should interact with, and the type of interaction for each program state (controlled-$\Zgate$, or controlled-$\Xgate$, or controlled-$\Ygate$), and where to insert the $\G$ gates and $\tgate$ gates between uses of the program states (including the information about which qubit any $\G$ gate or $\tgate$ gate acts on). Bob keeps track of the change in coefficients for the variables (in the linear polynomials to be evaluated by both parties) due to such program states.

4. For each $\tgate$ in the desired circuit, Alice does the following:\\
(1) Alice does the fixed $\G$ gates before a $\tgate$ gate, interspersed with some gadgets for single-qubit Clifford gates: for each program state corresponding to \emph{logical} gates before the current $\tgate$ gate, she does a controlled-$\Zgate$ (or controlled-$\Xgate$, controlled-$\Ygate$) gate on qubits $aq$, where $a$ is the received qubit and $q$ is a data qubit.\\
(2) Alice measures all such qubits $a$ in the $\{\ket{+},\ket{-}\}$ basis. She sends Bob all available measurement outcomes.\\
(3) Bob determines the coefficients in a linear polynomial according to Alice's messages and her previous messages, together with his knowledge about the program states. Alice and Bob perform the subprocedure for evaluating this linear polynomial for the qubit that a $\tgate$ gate acts on. Alice applies the $\tgate$ gate on such qubit, and, according to the output of the corresponding polynomial, performs the $\pgate^\dag$ gate on the same qubit.

5. For the last part of Clifford gates after the last $\tgate$ gate in the desired circuit, Alice does the following:\\
(1) Alice does the fixed Clifford gates including the $\G$ gates, interspersed with the gadget for some single-qubit Clifford gates: for each program state, she does a controlled-$\Zgate$ (or controlled-$\Xgate$, controlled-$\Ygate$) gate on qubits $aq$, where $a$ is the received qubit and $q$ is a data qubit.\\
(2) Alice measures all such qubits $a$ in the $\{\ket{+},\ket{-}\}$ basis. She sends Bob all available measurement outcomes.\\
(3)  Bob determines the coefficients in some linear polynomials according to Alice's messages and her previous messages, together with his knowledge about the program states. Alice and Bob perform the subprocedures for evaluating two linear polynomials for each output qubit. Alice applies $\Xgate^j \Zgate^k$ on each output qubit, where $j$ and $k$ are the output of the corresponding polynomials.\\
\indent This completes the protocol.\\

Note that in the step 1, Bob uses local measurement with the help of entanglement to prepare the program states on Alice's side, rather than direct sending of the program states, since any failure in preparation of the entangled pairs can be remedied by preparing a new pair, while direct sending of the program states may have some rate of failure which is not acceptable. Another remark is on the use of the three types of controlled-Pauli gates. In principle, controlled-$\Zgate$ gate together with some fixed Clifford gates suffice for replacing the role of all three types of controlled-Pauli gates with the fixed Clifford gates, e.g. in ancilla-driven blind quantum computation \cite{skm13}. But the use of three types of controlled-Pauli gates here partially reduces the need for fixed types of gates on Alice's side, while not leaking much information about the circuit to Alice. Anyway, the sequence of Clifford gates would form an almost random Clifford circuit, if the number of program states used is large, no matter they are for one or three types of controlled-Pauli gates.

The argument for the correctness of the scheme consists of two parts: one is that the program states and associated operations indeed perform the desired Clifford gates. The other is that the key-update rules for $\tgate$ gates are correct. Both follow from elementary calculations.

We introduce two classes of protocols to be used for the subprocedure of evaluating classical linear polynomials. The overall scheme is interactive under both classes of protocols. For techniques of turning the scheme into a constant-round scheme, see the use of a teleportation gadget in the constant-round QHE scheme in \cite{Yu19}. Such type of gadget was introduced in \cite{Dulek16}.

The first class of protocols is completely classical. Alice sends $k$ variables (mask bits) to Bob in plaintext, and hides all remaining mask bits ($2n-k$ in the case of general quantum input, and $n-k$ in the case of classical or product real input). The choices of the sent variables are the same among all linear polynomials. Bob tells Alice a linear polynomial containing only the variables that are unknown to him. The generic form of the original polynomial is $y=(c+\sum_{j=1}^{m} a_i x_i) \mod 2$, but since some $x_i$ are known to Bob, the number of variables appearing in Bob's polynomial becomes fewer. The constant term in Bob's polynomial absorbs his coefficients for those variables known to him.

The second class is the protocol for evaluating linear polynomials in \cite{Yu19}. It contains one of a few different quantum check-based protocols for generating initial classical correlations (called ``one-time tables'') in \cite{Yu19}. The one-time tables are generated prior to the main computation. Some checks are performed by one or both parties to prevent the opposite party from cheating. So the protocols may abort when some party insists on cheating. If the party being checked indeed wants to perform the computation, he (she) should almost not cheat, as cheating does not bring much benefit in learning the opposite party's data, since the checks are before the main computation. The protocol for evaluating linear polynomials in \cite{Yu19} has distributed output (the XOR of two remote bits), while the use here has output on Alice's side only, so we need Bob to send Alice one bit at the end, which partially affects Bob's privacy. (Such partial leakage is mitigated by possible recompilations of the circuit in the overall interactive QHE scheme). With that minor point set aside, our interactive QHE scheme under the second class of protocols has asymptotic information-theoretic data privacy and circuit privacy, under some choice of the check-based protocol for generating one-time tables. But the circuit privacy is slightly worse than that of the interactive scheme in \cite{Yu19}, which uses a so-called garden-hose gadget for performing a possible $\pgate^\dag$ correction after a $\tgate$ gate. The current scheme is easier to implement because it does not need to use such gadget, while the scheme in \cite{Yu19} cannot get rid of it because the location of the main computation is on Bob's side.

In the following we discuss the security of Scheme 1 under the first class of protocols, i.e. the partly-secure classical protocols. We first discuss the case of general quantum input. The input data is partially secure in Scheme 1. Some of the input qubits are protected by Pauli masks. Thus, even if the outcomes of measurements on the program registers are sent to Bob, he cannot find out any information about the input state on these qubits. Some qubits may have some of the Pauli masks known to Bob, thus he may measure his qubit in his entangled pair in some suitable basis, and together with the measurement outcome on Alice's program register, he can get some information about the data. Such measurement by Bob may affect the correctness of computation. In the case of classical input, Bob's similar operation can be regarded as copying of the classical input, and it does not affect the correctness. The case of classical or product real input would have better tradeoff between data privacy and circuit privacy, since Bob has much fewer coefficients to tell Alice in each polynomial, for the same value of $k$.

Under the first class of protocols for the subprocedure, the circuit in Scheme 1 is partially secure, and the degree of circuit privacy depends on the value of $k$ and the number of $\tgate$ gates in the desired circuit: a larger value of $k$ and more $\tgate$ gates imply better circuit privacy. By sending Alice some polynomials, Bob tells Alice some information about his program. But since each polynomial has $k$ terms missing (because Alice tells Bob the values of $k$ variables), Alice lacks knowledge about this part of the program. The amount of lack of information is about $k-1$ bits for each polynomial, since Bob also sends Alice the constant term in the polynomial, which carries one bit or less (the case of less than one bit may happen when the variables told to Bob are all zero, since then Bob's constant term must be zero). Such lack of information would accumulate across different polynomials (each corresponding to a $\tgate$ gate), so that Alice lacks a significant part of the knowledge about the overall program. Here, we have ignored the asymptotically vanishing amount of information about the circuit learnable through the sequence of controlled-Pauli gates told to Alice. Note that in each polynomial, Alice's lack of knowledge is limited to some fixed terms. This may mean that the part of circuit on some subset of the input qubits not related to these terms may be known quite well by Alice, provided that these qubits do not interact with the remaining qubits very often.

\section{Experimental demonstration}\label{sec3}

\begin{figure*}[ht]
\centering
\includegraphics[scale=1.12]{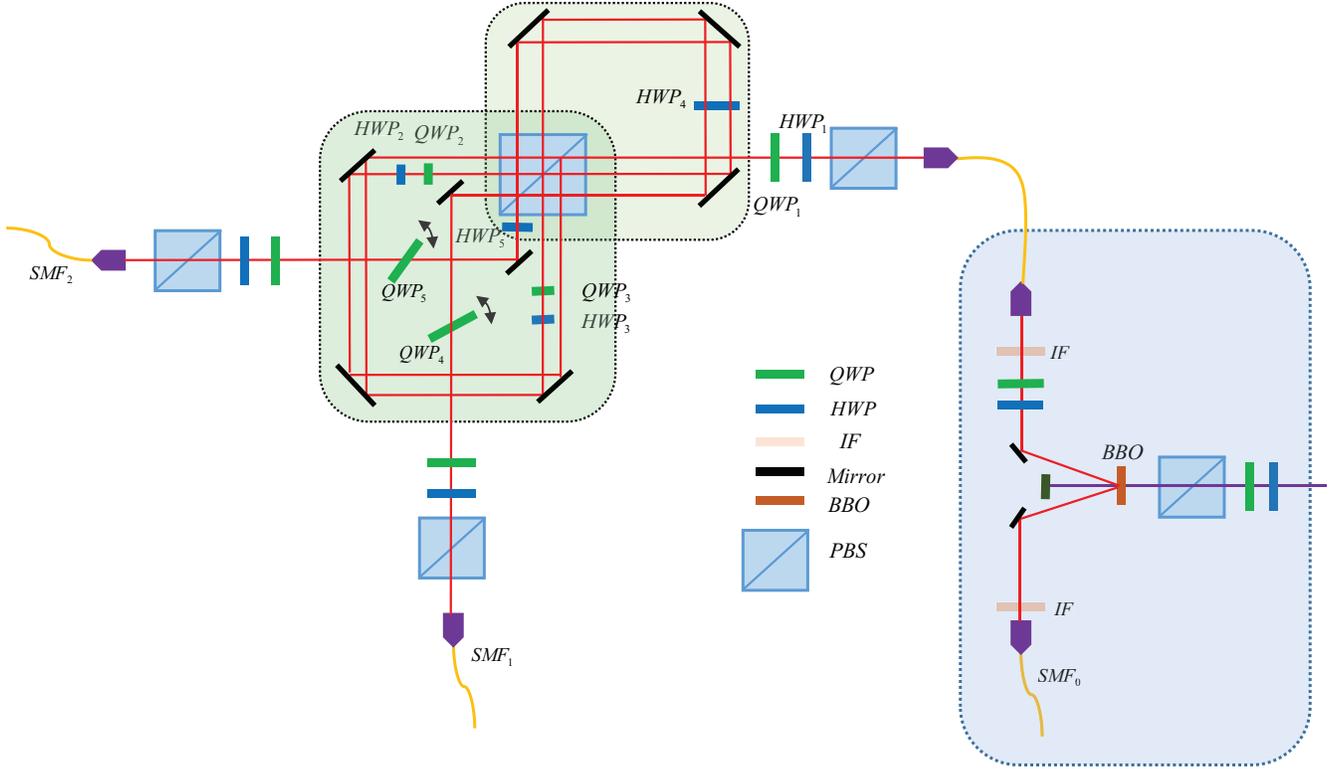}
\caption{The optical setup for demonstrating some key steps in the scheme. The HWP$_1$ and QWP$_1$ prepare a polarization state, and after the photon reaches the larger PBS at the middle, the prepared state becomes a path state, representing the controlling qubit in the controlled-Pauli gate to be performed. The HWP$_2$, QWP$_2$, HWP$_3$ and QWP$_3$ are for preparing polarization states dependent on the path, representing the results of the identity gate or Pauli gate acting on some initial polarization state. The HWP$_4$ and HWP$_5$ in the upper block are set to $22.5^\circ$ and $45^\circ$, respectively. The upper block serves to perform a Hadamard gate on the path state. The QWP$_4$ and QWP$_5$ are set to $0^\circ$ but tilted for phase compensation.}
\label{fig2}
\end{figure*}

We have performed a quantum optics experiment to demonstrate some steps in the outer part of the scheme: the state preparation of a data qubit and a program register (qubit), the controlled-Pauli gate on these qubits, the measurement on the program register after the gate, and state tomography on the output data qubit. The inner part, which is the subprocedure for evaluating classical linear polynomials, can be implemented independently, since it is connected to the outer part of the scheme classically. Such inner part can be implemented by known protocols or by those developed in the future. In our experiment, the controlled-$\Zgate$, controlled-$\Xgate$ and controlled-$\Ygate$ gates are performed using a single photon's path and polarization degrees of freedom. The photon's path is the controlling qubit (the program register) and the polarization is the controlled qubit. For ease of implementation, we assume directly the existence of a qubit as the program register as in Fig.~\ref{fig1}, instead of implementing it through measuring one qubit in an entangled pair of qubits as in Scheme 1. The optical circuit is illustrated in Fig.~\ref{fig2}.

In the setup, a pair of photons is generated by type-I spontaneous parametric down conversion in a 3-mm-thick nonlinear beta-barium borate (BBO) crystal pumped by a 100mW diode laser (centered at 404nm). The circuit first prepares a polarization state on one photon, while the other photon is used as the trigger. After passing through the larger PBS in the middle, the two branches go to separate paths, this means that the prior polarization state is actually used as a path state. In each of the two paths, the polarization state is known ($\ket{H}$ or $\ket{V}$), and we let the photon pass through some combination of HWP and QWP to get some polarization state dependent on the path. This effectively implements the controlled-$\sigma_j$ gate in Fig.~\ref{fig1}. Then the photon passes through the PBS again to enter the upper loop. This loop takes clue from the optical circuit in Fig. 2(b) of \cite{Liu16}. It consists of a PBS and some mirrors and HWPs (the HWP$_4$ and HWP$_5$ in Fig.~\ref{fig2}), and serves as an effective 50:50 beam splitter. The effective beam splitter performs a Hadamard gate on the path states. This implements the $\hgate$ gate in Fig.~\ref{fig1}. After the photon comes out of this loop (including passing through HWP$_5$ on one output path), it goes into one of the two output ports in the setup. The two path states at the end correspond to outcomes of the measurement on the lower qubit in Fig.~\ref{fig1}. In each run of the experiment, we only measure one polarization state in one output port. The choice of the port corresponds to choosing one outcome of the X-basis measurement on the ancilla qubit (in this notation, the $\hgate$ gate on the ancilla is regarded as part of the measurement). Under a fixed choice of the port, the choice of the polarization state to be detected is done by the QWP and HWP placed just before the detector.

Some separate testing shows that the interference visibility of the green (upper) loop and the grey (lower) loop are about 0.9983 and 0.9973, respectively.

For each combination of the input state of the controlling qubit, and the gate to be performed, we do five repeated periods of measurements. The duration of each measurement period is 20 seconds. We measure four polarization states in the selected output port, and reconstruct the density matrix using a program based on the maximum likelihood estimation method in \cite{JKM01}. For various input states of the controlling and the controlled qubits, and the chosen outcome for the measurement on the controlling qubit, we calculate the fidelity of the experimentally found density matrix with the ideal one. The fidelity between two pure states $\ket{\psi}$ and $\ket{\eta}$ is defined as $\vert\langle\psi\ket{\eta}\vert^2$. The fidelities for some parameterized sets of input states are listed in the figures below. Figure~\ref{fig3} shows fidelities between the theoretical and experimental output one-qubit states for the case of controlled-$Z$ gate acting on input state $\ket{1}\ox(\cos\theta\ket{0}+\sin\theta\ket{1})$ where $\theta$ is a real parameter, and subsequently measuring the controlling qubit with the outcome corresponding to $\ket{-}$. The average fidelities for other input state combinations are also mostly above 0.99.

The other figures illustrate the correctness of the results from another perspective. The left figure in Fig.~\ref{fig4} shows the theoretical curve and experimental result for the \emph{squared overlap} between the output state of the target qubit and the reference state $\ket{\phi_1}=\cos(\frac{\pi}{8})\ket{0}+e^{-i 4\pi/5}\sin(\frac{\pi}{8})\ket{1}$, for the case that the controlled-$Z$ gate acts on input state $\ket{1}\ket{\psi}$, with measurement outcome of the controlling qubit corresponding to $\ket{-}$ (same below). The squared overlap of two pure states has the same definition as the fidelity, and we use this notion to emphasize that it is not intended to be near $1$. The $\ket{\psi}$ here is not limited to states located on a circle on the Bloch sphere: the states used here include some states of the form $\cos\theta\ket{0}+\sin\theta\ket{1}$, and some states of the form $\cos\theta\ket{0}+i\sin\theta\ket{1}$, and another generic pure qubit state. The right figure in Fig.~\ref{fig4} is for the same input states, and the reference output state is changed to $\ket{\phi_2}=\cos(-\frac{\pi}{3})\ket{0}+e^{i \pi/7}\sin(-\frac{\pi}{3})\ket{1}$. The left figure in Fig.~\ref{fig5} shows the theoretical curve and experimental result for the squared overlap between the output state of the target qubit and the reference state $\ket{\phi_1}$ (same as that in Fig.~\ref{fig4}), for the case that the controlled-$Z$ gate acts on input state $\ket{y+}\ket{\psi}$, where $\ket{\psi}$ is not limited to states located on  a circle on the Bloch sphere. The right figure in Fig.~\ref{fig5} are for the same input states, but with the reference output state changed to $\ket{\phi_2}$ (same as that in Fig.~\ref{fig4}).

\begin{figure}[ht]
\centering
\includegraphics[scale=0.5]{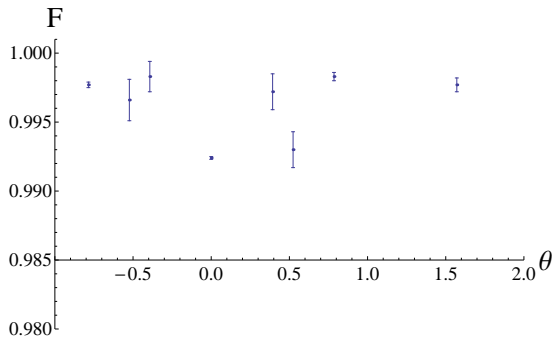}
\caption{Fidelities between the theoretical and experimental output one-qubit states in the case of a controlled-$Z$ gate acting on input state $\ket{1}\ox(\cos\theta\ket{0}+\sin\theta\ket{1})$ and subsequently measuring the controlling qubit with the outcome corresponding to $\ket{-}$ (the same outcome is assumed below). The error bars indicate fluctuations in the experimentally obtained fidelities over repeated runs of the experiment.}
\label{fig3}
\end{figure}

\begin{figure*}[htbp]
\centering
\subfigure{
\begin{minipage}[t]{0.5\linewidth}
\centering
\includegraphics[scale=0.55]{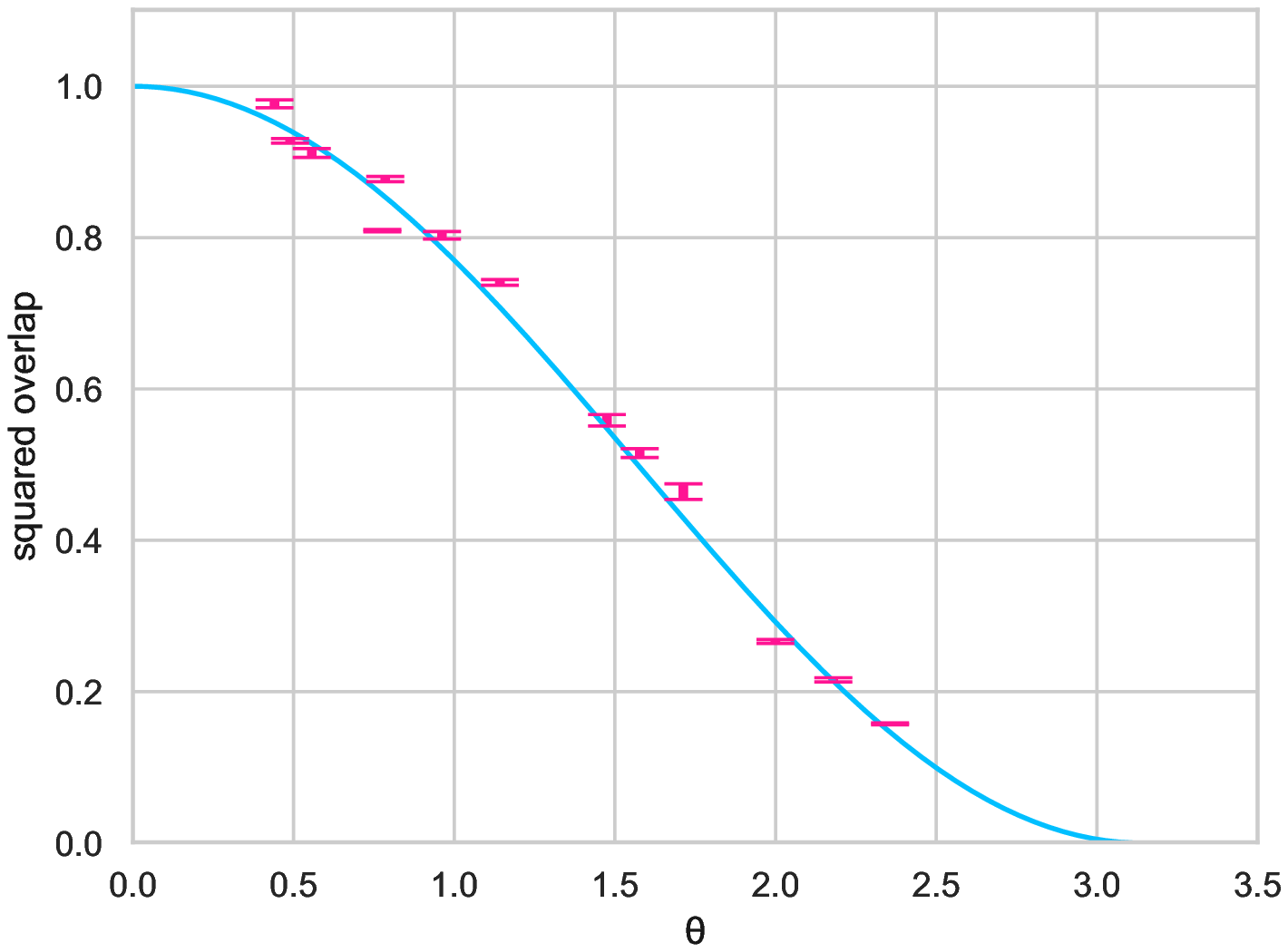}
\end{minipage}%
}%
\subfigure{
\begin{minipage}[t]{0.5\linewidth}
\centering
\includegraphics[scale=0.55]{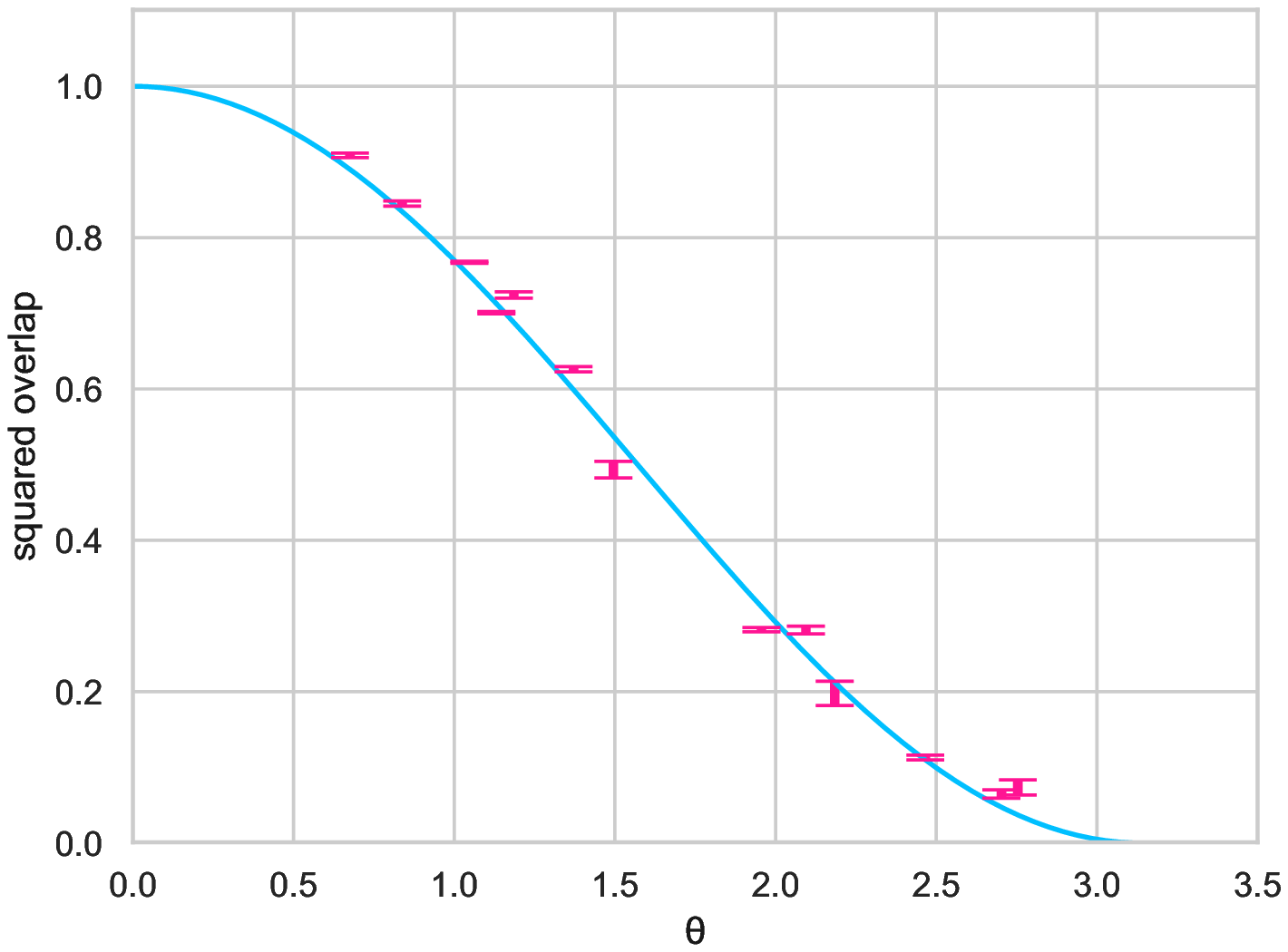}
\end{minipage}%
}%
\caption{Left: theoretical curve and experimental result for the squared overlap between the output state of the target qubit and a reference state $\ket{\phi_1}$, for the case that the controlled-$Z$ gate acts on $\ket{1}\ket{\psi}$, where $\ket{\psi}$ is not limited to states located on a circle on the Bloch sphere. The horizontal axis is for the angle with respect to the center of the Bloch sphere between the theoretical output state and the reference state $\ket{\phi_1}$. Right: for the same input states, but the reference state of the output is changed to $\ket{\phi_2}$.}
\label{fig4}
\end{figure*}

\begin{figure*}[htbp]
\centering
\subfigure{
\begin{minipage}[t]{0.5\linewidth}
\centering
\includegraphics[scale=0.55]{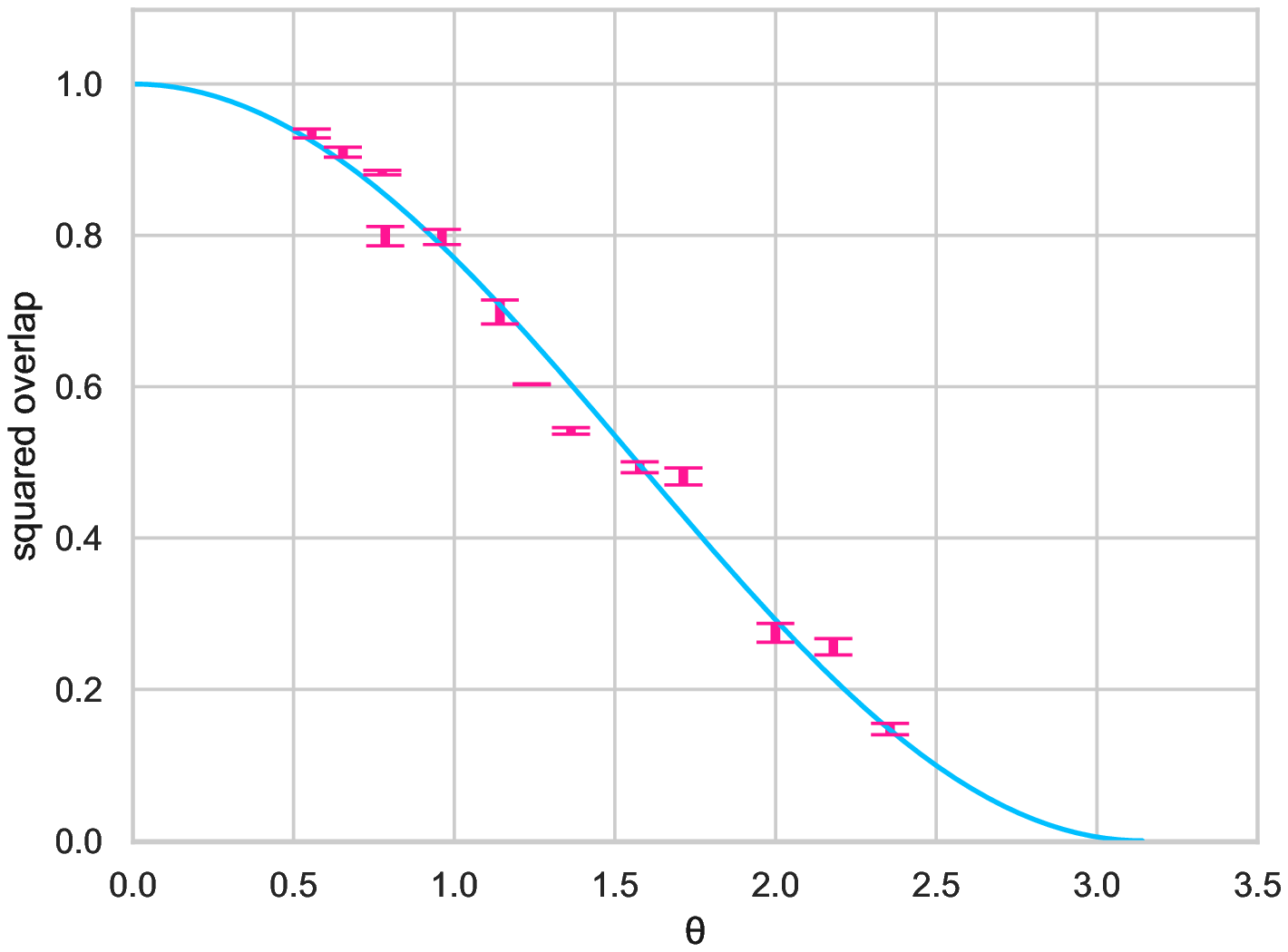}
\end{minipage}%
}%
\subfigure{
\begin{minipage}[t]{0.5\linewidth}
\centering
\includegraphics[scale=0.55]{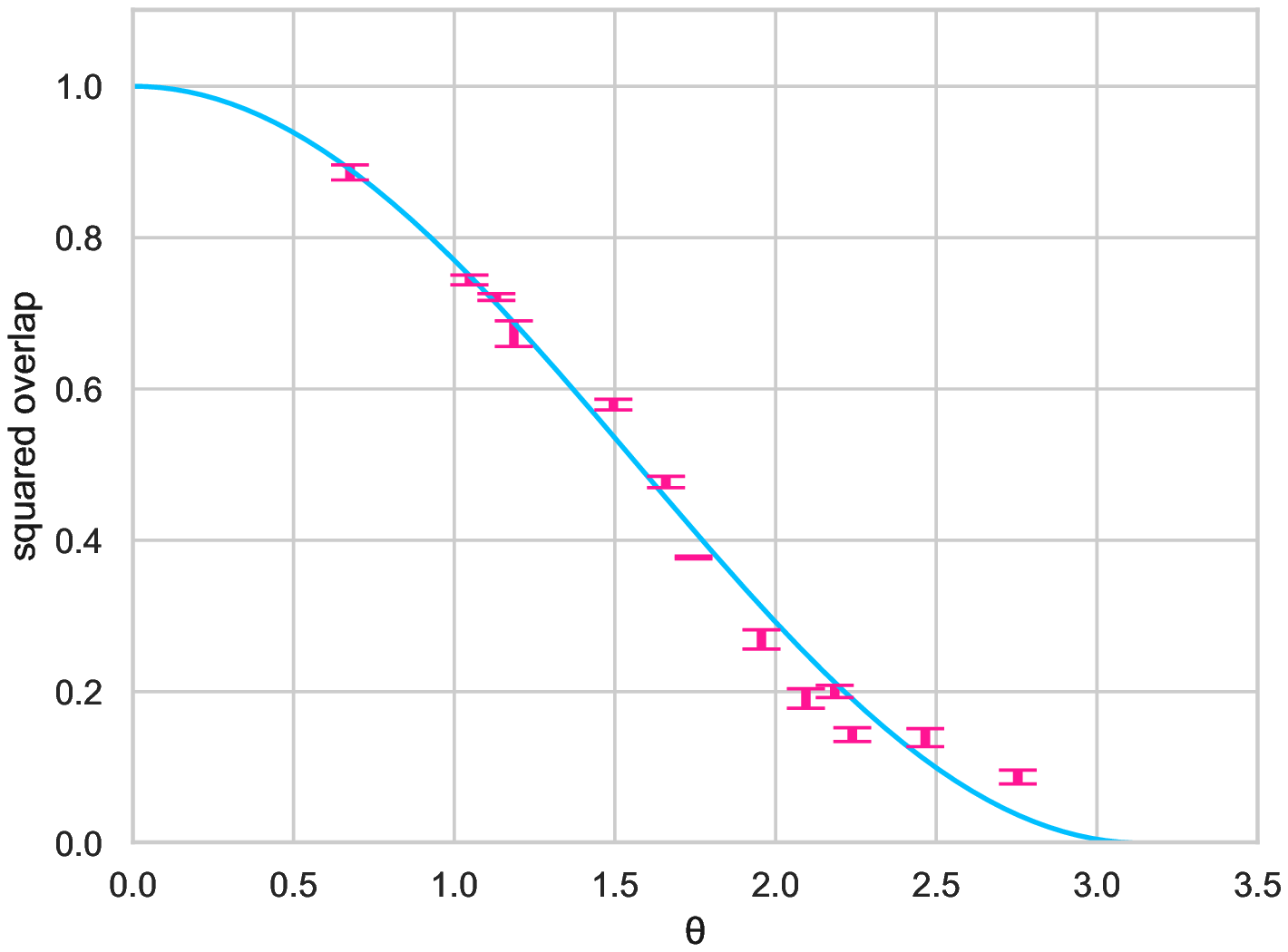}
\end{minipage}%
}%
\caption{Left: theoretical curve and experimental result for the squared overlap between the output state of the target qubit and a reference state $\ket{\phi_1}$, for the case that the controlled-$Z$ gate acts on $\ket{y+}\ket{\psi}$, where $\ket{\psi}$ is not limited to states located on a circle on the Bloch sphere. The horizontal axis is for the angle with respect to the center of the Bloch sphere between the theoretical output state and the reference state $\ket{\phi_1}$. Right: for the same input states, but the reference state of the output is changed to $\ket{\phi_2}$.}
\label{fig5}
\end{figure*}

\section{Discussion}\label{sec4}

Under the first class of protocols for evaluating classical linear polynomials, the extreme case of completely insecure data gives rise to an overall scheme similar to blind quantum computing (BQC) scheme, but with the output on a different party. In blind quantum computing, the client (usually called Alice, but to compare to our scheme, let us assume the client is Bob and the server is Alice) has both the data and the program but has limited quantum capabilities. The client delegates (a part of) the secret program to the server to let the program run on the client's private data (or a blank state), and the result is returned to the client in the end. Our scheme in the case of completely insecure data gives rise to a scheme for BQC if the final steps are modified so that Bob has the computation result instead of Alice. The revision required is that Alice teleports her output state to Bob before the final Pauli corrections, and Bob does not send her the information for the final Pauli corrections. The blindness is from that each program state is in a completely mixed qubit for the server. Such BQC scheme saves the number of rounds of communication compared to the scheme in \cite{bfk09} and some other existing schemes. On the other hand, without such modification, the scheme has the output on Alice's side, and a practical application is as follows: Bob serves as a data collecting party who effectively sends some data to Alice through the choices of the program states, and Alice does the main quantum computation, with the output known to herself. In such case, we no longer have the usual problem in BQC that the quantum computation needs to be verified, since Alice wants to get the result.

We remark on the use of program registers sent by the client Bob, and the associated necessity for Alice to encrypt her data locally (via the quantum one-time-pad in our scheme). In many BQC schemes, some qubit in mixed state (in the view of the server) is sent by the client to be used as a local program register on the server. Things are different in (interactive) QHE, in which the data and the program are initially on opposite parties. In schemes of (interactive) QHE, the use of a program register with Bob knowing its content is likely a weak point in security, since it is modelled by a mixed state, and we may think of two ways of purifying it, each has its own problem: (i) if a purifying third party is introduced, it may be an active party who may collude with some party or cheat; (ii) if instead a mixed state resource on the two parties (Alice and Bob) is used, there is no guarantee that Bob would not replace his part so that they share a pure entangled state with the same function in the protocol, and if Alice's data qubits are not properly encrypted, it is likely that Bob may measure his part of the entangled state at some stage of the protocol in some suitable basis, and together with the information about measurement outcomes sent by Alice, he may get some information about Alice's data. This issue does not cause a security problem in BQC, since in BQC the initial data is on the party who has the program. Our Scheme 1 avoids the problem above by the trick of letting Alice encrypt the data locally first.

The Scheme 1 contains a procedure of evaluating a classical linear polynomial of the form \eqref{eq:poly}. For such procedure, we have introduced a simple classical protocol with some privacy tradeoff as an example protocol. Such procedure could be replaced with any classical additive homomorphic encryption scheme encrypting bits, analogous to the situation in \cite{bj15}. One can use the relatively simple schemes in \cite{GM82,Paillier99} which are not post-quantum schemes; some post-quantum scheme may be used if the security against quantum computers is of concern.

\section{Conclusions}\label{sec5}

We have constructed a framework for (interactive) QHE scheme for general circuits, based on the universal circuit approach. A feature of the framework is the compatibility with blind quantum computation. We discussed two classes of example protocols for the subprocedure for evaluating a linear polynomial, and such subprocedure is replaceable by other protocols. With the first class of example protocols, which are fully classical, there is a tradeoff in the privacy of the data and the privacy of the circuit in the overall scheme, and the tradeoff is particularly good when the input is classical or a product real state. With the second class of example protocols which require repeated quantum operations on constant number of qubits, the overall scheme has asymptotic data privacy and asymptotic circuit privacy (the latter for large circuits only) under some weak assumptions about the two parties. We have demonstrated some key steps of the framework part of the scheme in a quantum optics experiment.

Apart from the implementation of the subprocedure, the framework has low requirement on Bob's quantum capabilities: he need only be able to prepare and send single-qubit states, or be able to make single-qubit measurements in one of two bases. We have mentioned in Sec.~\ref{sec2} that the framework may be made non-interactive by using the so-called garden-hose gadgets. Note that in a computationally-secure leveled QFHE scheme \cite{Mahadev17}, an almost classical client is also used, but the client provides the data rather than the circuit. As an alternative to the first type of protocols mentioned above, it is possible to use classical additive homomorphic encryption methods in evaluating the linear polynomials.

\smallskip
\section*{Acknowledgments}

This work was supported in part by the NKRDP of China (No. 2016YFA0301802), the National Natural Science Foundation of China (No. 11974096, No. U1705264, No. 11774076, No. 11775065, No. 61472114, No. 61972124, and No. 61672030), the Zhejiang Provincial Natural Science Foundation of China (No. LY17A050003, No. LY19F020019), the Research Foundation of Guangxi Key Laboratory of Cryptography and Information Security (No. GCIS201725), the Scientific Research Fund of Zhejiang Provincial Education Department (No. Y201737289, No. Y201737292), and the startup grant of Hangzhou Normal University.

\bibliographystyle{unsrt}
\bibliography{homo}

\end{document}